\documentclass[conference, a4paper]{IEEEtran}

\usepackage{algorithm,algorithmic}
\usepackage{amsmath,amssymb,mathrsfs}
\usepackage{array}
\usepackage{balance}
\usepackage{booktabs}
\usepackage{cite}
\usepackage{epsfig}
\usepackage{easyReview}
\usepackage{fancyhdr}
\usepackage{float}
\usepackage[nomain,acronym,automake,toc,nopostdot]{glossaries}
\usepackage{graphicx,color}
\usepackage[bookmarks=false]{hyperref}
\usepackage{mathtools}
\usepackage{multirow}
\usepackage{psfrag}
\usepackage{stfloats}
\usepackage{subfigure}
\usepackage{url}

\newtheorem{thm}{Theorem}
\newtheorem{rem}{Remark}

\makeglossaries
\newacronym[description=Additive white Gaussian noise]{awgn}{AWGN}{additive white Gaussian noise}
\newacronym[description= Alternating direction method of multipliers]{admm}{ADMM}{ alternating direction method of multipliers}
\newacronym[description=Approximate message passing]{amp}{AMP}{approximate message passing}
\newacronym[description={\em a posteriori} probability]{app}{APP}{{\em a posteriori} probability}
\newacronym[description=Base station]{bs}{BS}{base station}
\newacronym[description=Base station sleeping ]{bss}{BSS}{base station sleeping }
\newacronym[description=Belief propagation]{bp}{BP}{belief propagation}
\newacronym[description=Binary phase shift keying]{bpsk}{BPSK}{binary phase shift keying}
\newacronym[description=Bit error rate]{ber}{BER}{bit-error-rate}
\newacronym[description=Block error rate]{bler}{BLER}{block error rate}
\newacronym[description=Central limit theorem]{clt}{CLT}{central limit theorem}
\newacronym[description=Channel state information ]{csi}{CSI}{channel state information }
\newacronym[description=Closest vector problem]{cvp}{CVP}{closest vector problem}
\newacronym[description=Code division multiple access]{cdma}{CDMA}{code division multiple access}
\newacronym[description=Distributed linear data fusion]{dldf}{DLDF}{distributed linear data fusion}
\newacronym[description=European Cooperation in Science and Technology]{cost}{COST}{Cooperation in Science and Technology}
\newacronym[description=Coordinated multi-point ]{CoMP}{CoMP}{coordinated multi-point }
\newacronym[description=Correlation-based stochastic model]{cbsm}{CBSM}{correlation-based stochastic model}
\newacronym[description=Cumulative distribution function]{cdf}{CDF}{cumulative distribution function}
\newacronym[description=Degrees of freedom]{dof}{DoF}{degrees of freedom}
\newacronym[description=Element-based lattice reduction]{elr}{ELR}{element-based lattice reduction}
\newacronym[description=Extremely large aperture array]{elaa}{ELAA}{extremely large aperture array}
\newacronym[description=Fifth-generation]{5g}{5G}{fifth-generation}
\newacronym[description=Fixed-complexity sphere decoder]{fcsd}{FCSD}{fixed-complexity sphere decoder}
\newacronym[description=Forward error corrrection]{fec}{FEC}{forward error correction}
\newacronym[description=Free space path loss]{fspl}{FSPL}{free space path loss}
\newacronym[description=Gauss-Seidel]{gs}{GS}{Gauss-Seidel}
\newacronym[description=Global system for mobile communication]{gsm}{GSM}{global system for mobile communication}
\newacronym[description=Geometry-based stochastic model]{gbsm}{GBSM}{geometry-based stochastic model}
\newacronym[description=Hermite-Korkin-Zolotarev]{hkz}{HKZ}{Hermite-Korkin-Zolotarev}
\newacronym[description=Independent and identically distributed]{iid}{i.i.d.}{independent and identically distributed}
\newacronym[description=Independent and non-identically distributed]{ind}{i.n.d.}{independent and non-identically distributed}
\newacronym[description=Integer least-squares]{ils}{ILS}{integer least-squares}
\newacronym[description=Iterative discrete estimation]{ide}{IDE}{iterative discrete estimation}
\newacronym[description=Iterative discrete estimation 2]{ide2}{IDE2}{iterative discrete estimation 2}
\newacronym[description=International Telecommunication Union Radiocommunication Sector ]{itu-r}{ITU-R}{International Telecommunication Union Radiocommunication Sector}
\newacronym[description=Large system behaviour]{lsb}{LSB}{large system behaviour}
\newacronym[description=Lattice reduction]{lr}{LR}{lattice reduction}
\newacronym[description=Lenstra-Lenstra-Lov\'{a}sz]{lll}{LLL}{Lenstra-Lenstra-Lov\'{a}sz}
\newacronym[description=Likelihood ascent search]{las}{LAS}{likelihood ascent search}
\newacronym[description=Line-of-slight]{los}{LoS}{line-of-slight}
\newacronym[description=List sphere decoder]{lsd}{LSD}{list sphere decoder}
\newacronym[description=Linear minimum mean square error]{lmmse}{LMMSE}{linear minimum mean square error}
\newacronym[description=Log-likelihood ratio]{llr}{LLR}{log-likelihood ratio}
\newacronym[description=Long-term evolution ]{lte}{LTE}{long-term evolution}
\newacronym[description=Low density parity check]{ldpc}{LDPC}{low density parity check}
\newacronym[description=Massive machine type communications]{mmtc}{mMTC}{massive machine type communications}
\newacronym[description=Maximum {\em a posteriori}]{map}{MAP}{maximum {\em a posteriori}}
\newacronym[description=Maximum-likelihood]{ml}{ML}{maximum-likelihood}
\newacronym[description=Maximum-likelihood sequence detection]{mlsd}{MLSD}{maximum-likelihood sequence detection}
\newacronym[description=Maximum ratio combining]{mrc}{MRC}{maximum ratio combining}
\newacronym[description=Multiple-input multiple-output]{mimo}{MIMO}{multiple-input multiple-output}
\newacronym[description=Massive multiple-input multiple-output]{mmimo}{mMIMO}{massive multiple-input multiple-output}
\newacronym[description=Matched filter]{mf}{MF}{matched filter}
\newacronym[description=Matched-filter bound]{mfb}{MFB}{matched-filter bound}
\newacronym[description=Mean square error]{mse}{MSE}{mean square error}
\newacronym[description=Minimum mean square error]{mmse}{LMMSE}{minimum mean square error}
\newacronym[description=Mobile and wireless communications Enablers for the Twenty-twenty Information ]{metis}{METIS}{Mobile and wireless communications Enablers for the Twenty-twenty Information}
\newacronym[description=Non-line-of-sight]{nlos}{NLoS}{non-LoS}
\newacronym[description=Non-deterministic polynomial-time hard]{nphard}{NP-hard}{non-deterministic polynomial-time hard}
\newacronym[description=One dimensional]{1d}{1-D}{one dimensional}
\newacronym[description=Orthogonality defect]{od}{OD}{orthogonality defect}
\newacronym[description=Pairwise error probability]{pep}{PEP}{pairwise error probability}
\newacronym[description=Parallel interference cancellation]{pic}{PIC}{parallel interference cancellation}
\newacronym[description=Preconditioning for iterative mMIMO detection]{pid}{PID}{preconditioning for iterative mMIMO detection}
\newacronym[description=Probabilistic data association]{pda}{PDA}{probabilistic data association}
\newacronym[description=Probability distribution function]{pdf}{PDF}{probability density function}
\newacronym[description=Probability mass function]{pmf}{PMF}{probability mass function}
\newacronym[description=Quadrature amplitude modulation]{qam}{QAM}{quadrature amplitude modulation}
\newacronym[description=Quadrature phase shift keying]{qpsk}{QPSK}{quadrature phase shift keying}
\newacronym[description=Receiver-side channel state information]{rcsi}{R-CSI}{receiver-side channel state information}
\newacronym[description=Received signal strength]{rss}{RSS}{received signal strength}
\newacronym[description=Semidefinite relaxation]{sdr}{SDR}{semidefinite relaxation}
\newacronym[description=Seysen's algorithm]{sa}{SA}{Seysen's algorithm}
\newacronym[description=Signal-to-interference-plus-noise ratio]{sinr}{SINR}{signal-to-interference-plus-noise ratio}
\newacronym[description=Signal to interference ratio]{sir}{SIR}{signal to interference ratio }
\newacronym[description=Signal-to-noise ratio]{snr}{SNR}{signal-to-noise ratio}
\newacronym[description=Single antenna interference cancellation]{saic}{SAIC}{single antenna interference cancellation}
\newacronym[description=Single input single output]{siso}{SISO}{single input single output}
\newacronym[description=Singular value decomposition ]{svd}{SVD}{singular value decomposition }
\newacronym[description=Sixth-generation mobile networks]{6g}{6G}{sixth-generation}
\newacronym[description=Sphere decoder]{sd}{SD}{sphere decoder}
\newacronym[description=Space-time codes]{stc}{STC}{space-time codes}
\newacronym[description=State-of-the-art]{sota}{SoTA}{state-of-the-art}
\newacronym[description=Successive interference cancellation]{sic}{SIC}{succesive interference cancellation}
\newacronym[description=Symbol error rate]{ser}{SER}{symbol error rate}
\newacronym[description=Symmetric successive over-relaxation]{ssor}{SSOR}{symmetric successive over-relaxation}
\newacronym[description=Tabu search]{ts}{TS}{tabu search}
\newacronym[description=Three-dimensional]{3d}{3-D}{three-dimensional}
\newacronym[description=The 3rd Generation Partnership Project]{3gpp}{3GPP}{the 3rd Generation Partnership Project}
\newacronym[description=Two-dimensional]{2d}{2-D}{two-dimensional}
\newacronym[description=Uniform linear array]{ula}{ULA}{uniform linear array}
\newacronym[description=Urban micro]{umi}{UMi}{urban micro}
\newacronym[description=User equipment]{ue}{UE}{user equipment}
\newacronym[description=Vector error rate]{ver}{VER}{vector error rate}
\newacronym[description=Vertical Bell Labs layered space-time]{vblast}{V-BLAST}{vertical Bell Labs layered space-time}
\newacronym[description=Visibility region]{vr}{VR}{visibility region}
\newacronym[description=Widely linear]{wl}{WL}{widely linear}
\newacronym[description=Widely linear zero forcing]{wlzf}{WLZF}{widely linear zero forcing}
\newacronym[description=Wide-sense stationary uncorrelated scattering]{wssus}{WSSUS}{wide-sense stationary uncorrelated scattering}
\newacronym[description=Wireless World Initiative New Ratio]{winner}{WINNER}{Wireless World Initiative New Ratio}
\newacronym[description=Extra-large multiple-input multiple-output]{xlmimo}{XL-MIMO}{extra-large multiple-input multiple-output}
\newacronym[description=Zero-forcing]{zf}{ZF}{zero-forcing}
\newacronym[description=Zero mean complex circularly symmetric]{zmccs}{ZMCCS}{zero mean complex circularly symmetric}

\definecolor{sblue}{RGB}{0,51,120}
\definecolor{sred}{RGB}{200,51,130}

\newcommand{\figref}[1]{Fig. \ref{#1}}
\newcommand{\tabref}[1]{TABLE \ref{#1}}
\newcommand{\secref}[1]{Section \ref{#1}}
\renewcommand{\eqref}[1]{(\ref{#1})}

\ifCLASSINFOpdf
\else
\fi

\begin{document}
\title{Alternative Normalized-Preconditioning for Scalable Iterative Large-MIMO Detection}
\author{Jiuyu Liu, Yi Ma, and Rahim Tafazolli\\
	{\small 5GIC and 6GIC, Institute for Communication Systems, University of Surrey, Guildford, UK, GU2 7XH}\\
	{\small Emails: (jiuyu.liu, y.ma, r.tafazolli)@surrey.ac.uk}}
\markboth{}%
{}

\maketitle

\begin{abstract}
Signal detection in large multiple-input multiple-output (large-MIMO) systems presents greater challenges compared to conventional massive-MIMO for two primary reasons.
First, large-MIMO systems lack favorable propagation conditions as they do not require a substantially greater number of service antennas relative to user antennas.
Second, the wireless channel may exhibit spatial non-stationarity when an extremely large aperture array (ELAA) is deployed in a large-MIMO system.
In this paper, we propose a scalable iterative large-MIMO detector named ANPID, which simultaneously delivers  {\it 1)} close to maximum-likelihood detection performance, {\it 2)} low computational-complexity (i.e., square-order of transmit antennas), {\it 3)} fast convergence, and {\it 4)} robustness to the spatial non-stationarity in ELAA channels.
ANPID incorporates a damping demodulation step into stationary iterative (SI) methods and alternates between two distinct demodulated SI methods.
Simulation results demonstrate that ANPID fulfills all the four features concurrently and outperforms existing low-complexity MIMO detectors, especially in highly-loaded large-MIMO systems.
\end{abstract}

\section{Introduction}\label{sec1}
In forthcoming wireless communication paradigms, large multiple-input multiple-output (large-MIMO) system will play an important role for serving a large number of user devices \cite{Cui2023}.
For instance, within the framework of smart city infrastructures, there will be hundreds or even thousands internet-of-things (IoT) devices.
Therefore, compared to conventional massive-MIMO systems, the receiver design for large-MIMO systems presents unique challenges for two main reasons.
First, large-MIMO systems do not require a substantially larger number of service antennas compared to user antennas, which implies that the favorable propagation assumption cannot be ensured \cite{BJORNSON20193}.
Second, when an extremely large aperture array (ELAA) is incorporated into the system, the wireless channel can become spatially non-stationary \cite{Liu2021}.
Developing a scalable large-MIMO detector necessitates addressing the following requirements simultaneously: {\it 1)} low computational-complexity, {\it 2)} near-optimal performance, {\it 3)} fast convergence, and {\it 4)} robustness to the channel spatial non-stationarity \cite{Yang2015, Albreem2019}.
Current MIMO detectors can be categorized as either linear or nonlinear, based on their detection performance.

Linear MIMO detectors, such as minimum mean square error (LMMSE) and zero-forcing (ZF), are two of the most well-known detectors in the literature \cite{Albreem2019}. 
These detectors can achieve near-optimal performance when the MIMO channel is well-conditioned \cite{Chen2018}.
However, they both require a channel matrix inverse with cubic-order complexity of the user-antenna number, limiting their scalability.
To address this issue, researchers have proposed several types of iterative methods, mainly including stationary iterative (SI) methods, gradient descent methods, quasi-Newton methods, and belief propagation.
Although these methods have scalable complexities, their detection performances become too sub-optimal when the system load is high since they can only converge to the performance of ZF/LMMSE \cite{Liu2023a}.

Nonlinear MIMO detectors comprise various search-based approaches \cite{Yang2015}, such as maximum likelihood sequence detection (MLSD), likelihood ascent search, sphere decoding, $K$-best search, tabu search, and etc. 
While these detectors can potentially achieve (near-) MLSD performance, their complexities render them impractical, particularly for a large number of user antennas or high modulation orders \cite{Albreem2019}.
Recently, several MIMO detectors based on alternating direction method of multipliers (ADMM) framework have been proposed to solve the MLSD problem subject to constraints \cite{Albreem2022,Shahabuddin2021,Tiba2021,Zhang2022,Chen2017}, such as penalty-sharing ADMM (PS-ADMM).
In comparison to linear detectors, ADMM-based methods provide significant performance gains in symmetric large-MIMO systems with stationary channels.
However, these methods all involve the inversion or decomposition of a Gram matrix, which limits their scalability with user antennas.

In this paper, we propose a nonlinear MIMO detector called Alternative Normalized-Preconditioning for Iterative large-MIMO Detection (ANPID).
Drawing inspiration from ADMM-based methods, our approach involves performing demodulation of three SI methods at each iteration: Jacobi iteration, Gauss-Seidel (GS) method, and symmetric successive over-relaxation (SSOR).
Then, a damping step is developed to combine the estimation and demodulation vectors.
Although GS and SSOR converge faster than Jacobi iteration, the signal power is not normalized in their iterative processes, which is inconsistent with the demodulation step.
To address this issue, a well-designed diagonal matrix is proposed to normalized the signal power.
Finally, the normalized GS/SSOR and Jacobi methods are alternately employed to achieve both fast convergence and close-to-MLSD performance.

The complexity of ANPID is comparable to that of SI methods, featuring scalable complexity, i.e., square-order of user-antenna number, as the projection and damping steps entail negligible complexities.
Moreover, our simulation results demonstrate that ANPID achieves fast convergence and near-MLSD performance, even in highly-loaded and spatially-non-stationary large-MIMO systems.

\section{Signal Model, Preliminaries and Problem Statement}\label{sec2}
\subsection{Signal Model}
Let $M$ and $N$ denote the number of service and user antennas, respectively. The uplink signal model of large-MIMO systems can be represented as follows
\begin{equation}\label{eqn01}
	\mathbf{y} = \mathbf{H} \mathbf{x} + \mathbf{v},
\end{equation}
where $\mathbf{y} \in \mathbb{C}^{M \times 1}$ stands for the received signal vector, $\mathbf{H} \in \mathbb{C}^{M \times N}$ for the random channel matrix, $\mathbf{x} \in \mathbb{C}^{N \times 1}$ for the transmitted signal vector,
$\mathbf{v} \sim \mathcal{CN}(0,\sigma_v^2\mathbf{I}_M)$ for the additive white Gaussian noise (AWGN), and $\mathbf{I}_M$ for an $(M)\times(M)$ identity matrix.
Each element of $\mathbf{x}$ is assumed to be drawn from a modulation constellation $\mathcal{A}$ with equal probability, and fulfills: $\mathbb{E}\{\mathbf{x}\} = \mathbf{0}$; $\mathbb{E}\{\mathbf{x}\mathbf{x}^{H}\} = \sigma^{2}_x\mathbf{I}_{N}$.
The following two random distributions of $\mathbf{H}$ are considered in this paper:

\subsubsection{WSSUS channel} 
In conventional MIMO systems, the wireless channel is typically assumed to be wide-sense stationary uncorrelated scattering (WSSUS) \cite{Hochwald2004}, where each element of $\mathbf{H}$ obeys independent and identically distributed (i.i.d.) Rayleigh fading as follows
\begin{equation}
h_{m,n} \sim \mathcal{CN}\left(0,\sigma^2_h \right),\ \forall m, n,
\end{equation}
where $\sigma^2_h$ denote the variance of every channel element.

\subsubsection{ELAA channel} 
The wireless channel may become spatially non-stationary when an ELAA is deployed in an large-MIMO system.
It is more appropriate to use a spherical-wave model to describe the spatial non-stationarity as follows \cite{Liu2023}
\begin{equation}\label{eqn02}
	h_{m,n}^{\textsc{elaa}} = \varepsilon_{m,n} \left(\frac{\alpha}{d_{m,n}^{\beta}}\right) h_{m,n},
\end{equation}
where $\varepsilon_{m,n} \sim \mathcal{LN}(0,\sigma_s)$ stands for the log-normal distributed shadowing effects, $\alpha$ for the path-loss coefficients, $\beta$ for the path-loss exponents and $d_{m,n}$ for the distance between the $m^{th}$ service and $n^{th}$ user antennas.
The correlation of $\varepsilon_{m,n} $ is characterized by exponentially decaying window in \cite{Liu2021}.

\subsection{Preliminaries} \label{sec2b}
Denote $\mathbf{x}_{\textsc{ml}}$ as the decision of MLSD, it can be determined by solving the following integer least squares problem
\begin{equation}\label{eqn03}
	\mathbf{x}_{\textsc{ml}}=\underset{\mathbf{x}\in\mathcal{A}^{N}}{\arg \min}\ \|\mathbf{y} - \mathbf{H} \mathbf{x}\|^{2},
\end{equation} 
where $\|\cdot\|$ denotes the Euclidean norm.
However, it is computationally prohibitive to obtain $\mathbf{x}_{\textsc{ml}}$ for large value of $N$. 
As a lower-complexity alternative, several linear MIMO detectors (e.g., ZF and LMMSE) have been proposed. 
Their decision vectors, denoted by $\mathbf{x}_{\textsc{lin}}$, can be expressed as follows
\begin{equation}\label{eqn05}
	\mathbf{x}_{\textsc{lin}} = \Gamma(\mathbf{A}^{-1}\mathbf{b}),
\end{equation}
where $\mathbf{A} \triangleq \mathbf{H}^H\mathbf{H} + \rho \mathbf{I}_{N}$, $\mathbf{b} \triangleq \mathbf{H}^H \mathbf{y}$ and $\Gamma(\cdot)$ performs symbol-by-symbol decision.
Specifically, the detector in \eqref{eqn05} becomes ZF when the regularization factor $\rho = 0$, and LMMSE when $\rho = \sigma^{2}_{v}/\sigma^{2}_{s}$.
However, due to the inversion of $\mathbf{A}$, their complexities are still not scalable as $N$ increases.
SI methods are proposed to bypass the matrix inverse as follows
\begin{equation}\label{eqn07}
\mathbf{s}_{t} =\mathbf{s}_{t-1} + \mathbf{M}^{-1}(\mathbf{b} - \mathbf{A}\mathbf{s}_{t-1}),
\end{equation}
where $t \geq 1$ denotes the iteration index, $\mathbf{s}_{t}$ the $t^{th}$ estimation vector, and $\mathbf{M}$ is the preconditioning matrix to distinguish between different SI methods.
For examples, $\mathbf{M}_\text{Jacobi} = \mathbf{D}$, $\mathbf{M}_{\textsc{gs}} = \mathbf{D} + \mathbf{L}$ \cite{Zhang2021}, and $\mathbf{M}_{\textsc{ssor}} = \left(\mathbf{D+L}\right)\mathbf{D}^{-1}\left(\mathbf{D+L}\right)^H$ \cite{Xie2016}, where $\mathbf{D}$ and $\mathbf{L}$ are the diagonal and strict lower triangular parts of $\mathbf{A}$, respectively.

\subsection{Problem Statement}
In future large-MIMO systems, highly loaded situations will become increasingly common, and the wireless channels may exhibit spatially non-stationary.
In \secref{sec1} and \ref{sec2b}, existing MIMO detectors are discussed.
Linear detectors are only effective when $N \ll M$, while nonlinear detectors have high computational-complexity, i.e., $\mathcal{O}(N^3)$ or more.
To address these challenges, there is a pressing need for an advanced iterative large-MIMO detector that can satisfy the following key requirements simultaneously: \textit{1)} close-to-MLSD performance even in highly loaded systems, \textit{2)} low computational-complexity, i.e., $\mathcal{O}(N^2)$ or less, \textit{3)} fast convergence, and \textit{4)} robustness to channel spatial non-stationarity.

\section{The Development of ANPID Method}
In this section, we will provide a details of developing ANPID.
This involves integrating a damping demodulation step into SI methods, determining the optimal damping factor, designing normalized-SI techniques, and examining an alternative method on normalized-SI approaches.
Furthermore, pseudocode for the implementation of ANPID will be provided.

\subsection{Damping Demodulation for SI Methods}
Inspired by the demodulation step in ADMM-based methods, we propose incorporating a demodulation process after SI methods.
Additionally, a damping mechanism is introduced to combine the estimation and decision vectors as follows
\begin{subequations}\label{eqn08}
	\begin{align}
		\mathbf{s}_{t} &=\mathbf{d}_{t-1}+ \mathbf{M}^{-1}(\mathbf{b}-\mathbf{A}\mathbf{d}_{t-1}), \label{eqn08a}\\
		\mathbf{x}_{t} &=\Gamma \left(\mathbf{s}_{t}\right), \label{eqn08b}\\
		\mathbf{d}_{t} &=\omega_{t}\mathbf{d}_{t-1} + (1- \omega_{t}) \mathbf{x}_{t}, \label{eqn08c}   
	\end{align}
\end{subequations}
where $\mathbf{x}_{t}$, $\omega_{t}$ and $\mathbf{d}_{t}$ denote the decision vector, damping factor, and damping vector, respectively.
As with any iterative method, selecting an appropriate value of $\omega_{t}$ is crucial for achieving fast convergence and good detection performance.

\begin{thm} \label{thm01}
Given $\mathbf{H}$, $\mathbf{y}$, $\mathbf{x}_{t}$ and $\mathbf{d}_{t-1}$, the follow expression of $\omega_t$ minimizes the Euclidean distance between the received signal vector $\mathbf{y}$ and the damping vector $\mathbf{d}_{t}$ 
\begin{equation}\label{eqn15}
	\omega_{t} = \dfrac{\Re \left( \boldsymbol{\nu}_{t}^H\boldsymbol{\tau}_{t}\right) }{\|\boldsymbol{\nu}_{t}\|^2},
\end{equation}
where $\boldsymbol{\tau}_{t} = \mathbf{y}-\mathbf{H} \mathbf{x}_{t}$ and $\boldsymbol{\nu}_{t} = \mathbf{H} \mathbf{d}_{t-1} - \mathbf{H}\mathbf{x}_{t}$, and $\Re(\cdot)$ outputs the real part of the input vector.
\end{thm}

\begin{IEEEproof}
See Appendix \ref{app01}.
\end{IEEEproof}

\begin{figure}[t]
	\centering
	\includegraphics[width=7.8cm]{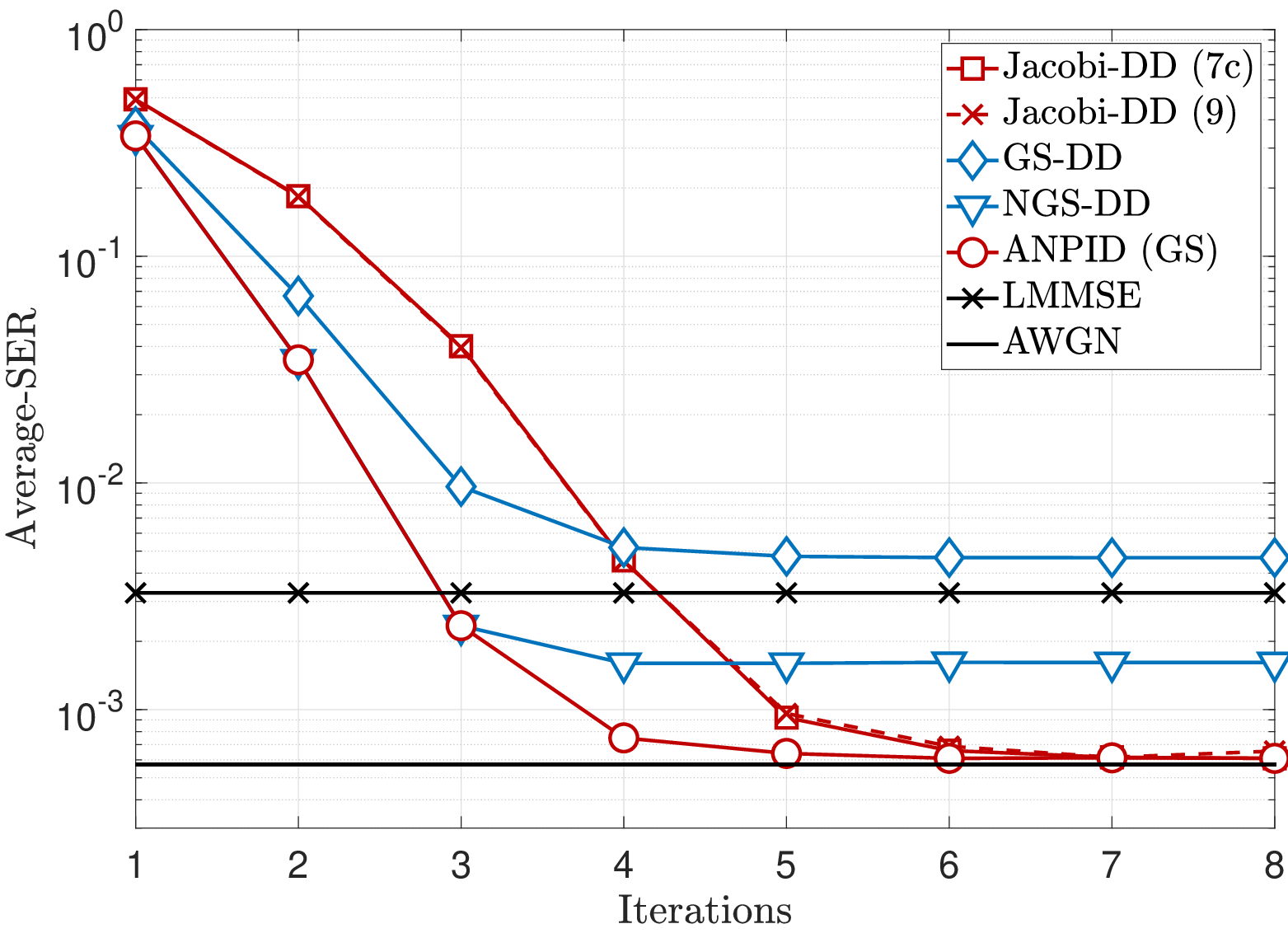}	
	\caption{SER versus iteration of different algorithms in i.i.d. Rayleigh fading channel. $M = 256$; $N = 64$; $16$ QAM; $\text{Es/No} = 18$ $\mathrm{dB}$.}
	\label{fig01}
	\vspace{-1em}
\end{figure}

\begin{rem}
The complexity of calculating ${\omega}_{t}$ using \eqref{eqn15} is $\mathcal{O}(2MN)$.
While this complexity is scalable, updating ${\omega}_{t}$ at every iteration is not necessary.
Our simulation results (see \figref{fig01}) indicate that setting $\omega^{\star} = {\omega}_{1}$ for all iterations achieves nearly the same convergence rate and symbol error rate (SER). Consequently, \eqref{eqn08c} can be replaced by 
\begin{equation}\label{eqn09010604}
	\mathbf{d}_{t} = {\omega}^{\star}\mathbf{d}_{t-1} + (1- {\omega}^{\star}) \mathbf{x}_{t},
\end{equation}
More specifically, $\omega^{\star}$ can be expressed as follows
\begin{equation}\label{eqn09}
	{\omega}^{\star} =  1- \dfrac{\Re \left(\mathbf{y}^{H}\mathbf{H} \mathbf{x}_{1}\right)}{ \|\mathbf{H}\mathbf{x}_{1}\|^2}, 
\end{equation}
which can be easily calculated by plugging $\mathbf{d}_{0} = \mathbf{0}$ into \eqref{eqn15}.
The SI methods that use \eqref{eqn09010604} as the damping step are referred to as Jacobi-DD, GS-DD, and SSOR-DD, respectively.
\end{rem}

In \figref{fig01}, the wireless channel is assumed to follow i.i.d. Rayleigh fading. 
The detection performance of Jacobi-DD is comparable to that of the AWGN-bound, suggesting that it can achieve close-to-MLSD performance as well.
However, GS-DD falls into a sub-optimal point with poor SER, despite its initially faster convergence compared to Jacobi-DD.
SSOR-DD also encounters this issue, but the results are not shown here due to space constraints.
This problem arises from the biased nature of \eqref{eqn08a} for GS/SSOR, as the signal power is not normalized before performing the demodulation in \eqref{eqn08b}.
Consequently, it is crucial to normalize the signal power in order to ensure consistency between these two steps.

\subsection{Normalized-SI Methods} \label{sec3b}
\begin{thm}\label{thm02}
Given $\mathbf{A} = \mathbf{H}^H\mathbf{H}$, the signal power for each data stream in SI-methods can be normalized to $1$ by replacing \eqref{eqn08a} as follows
\begin{equation}\label{eqn10}
	\mathbf{s}_{t}=\mathbf{d}_{t-1}+ (\mathbf{M}\mathbf{U})^{-1}(\mathbf{b}-\mathbf{A}\mathbf{d}_{t-1}),
\end{equation}
where $\mathbf{U}$ represents the diagonal part of $[\mathbf{M}^{-1}\mathbf{A}]$.
\end{thm}
\begin{IEEEproof}
See Appendix \ref{app02}.
\end{IEEEproof}

By employing \eqref{eqn10} instead of \eqref{eqn08a}, the proposed methods can be referred to as NGS/NSSOR.
It is worth noting that the signal power for the Jacobi iteration is already normalized to one, and inline with {\it Theorem \ref{thm01}}.
The performance comparison illustrated in \figref{fig01} demonstrates that NGS-DD outperforms GS-DD in terms of both convergence rate and SER performance. 
When comparing NGS-DD and Jacobi-DD, it becomes evident that NGS-DD has a faster convergence rate, while Jacobi-DD shows better SER performance. 
Consequently, the subsequent section aims to combine the advantages of both methods.

\subsection{ANPID Method with Pseudocode}
In this section, by alternating use NGS/NSSOR-DD and Jacobi-DD methods, two types of ANPID methods are proposed: ANPID (GS) and ANPID (SSOR).
The ANPID methods comprise two stages, specifically Stage A and Stage B, and their iterative process can be expressed as:
\begin{subequations}\label{eqn11}
	\begin{align}
		\mathbf{s}_{t} &=\mathbf{d}_{t-1}+\boldsymbol{\Theta}(\mathbf{b}-\mathbf{A}\mathbf{d}_{t-1}), \label{eqn11a}\\
		\mathbf{x}_{t} &=\Gamma\left(\mathbf{s}_{t}\right), \label{eqn11b}\\
		\mathbf{d}_{t} &=\zeta\mathbf{d}_{t-1}+(1 - \zeta)\mathbf{x}_{t}, \label{eqn11c}
	\end{align}
\end{subequations}
where $\boldsymbol{\Theta}$ and $\zeta$ represent the alternative preconditioning matrix and damping factor, respectively.
The value of $\boldsymbol{\Theta}$ and $\zeta$ will be alternated between Stage A and Stage B.
For instance, in the case of ANPID (GS) method, NGS-DD is utilized in Stage A for fast convergence, where $\boldsymbol{\Theta} = (\mathbf{M}_{\textsc{gs}}\mathbf{U})^{-1}$ and $\zeta_{\textsc{a}} = \omega_{\textsc{gs}}$.
Subsequently, in Stage B, Jacobi-DD is implemented for close-to-MLSD performance, where $\boldsymbol{\Theta} = \mathbf{D}^{-1}$ and $\zeta_{\textsc{b}} = \omega_{\textsc{Jac}}$.
The pseudocode of ANPID (GS) method is shown below:
\vspace{-0em}
\begin{algorithm}[h]
	{\small \renewcommand{\thealgorithm}{}
		\caption{\label{alg} ANPID (GS) Method} 
		\begin{algorithmic}[1]\label{alg01}
			\renewcommand{\algorithmicrequire}{\textbf{Input:}} 
			\REQUIRE~\\ $\mathbf{A}$; $\mathbf{b}$; $\mathbf{M}_{\textsc{gs}}$; $\mathbf{d}_0 = \mathbf{0}$; $T_{\textsc{a}}/T_{\textsc{b}}$: iterations of Stage A/B
			\renewcommand{\algorithmicrequire}{\textbf{Output:}} 
			\REQUIRE~\\${\mathbf{x}}_{t}$: the decision vector;
			\renewcommand{\algorithmicensure}{\textbf{START}}
			\ENSURE  
			\STATE {\bf let} $t = 1$, $\boldsymbol{\Theta} = (\mathbf{M}_{\textsc{gs}}\mathbf{U})^{-1}$; call \eqref{eqn11a} \eqref{eqn11b} to compute $\mathbf{x}_{1}$, ${\mathbf{z}}_{1}$; call \eqref{eqn09} to compute $\omega^{\star}_{\textsc{gs}}$ and $\omega^{\star}_{\text{Jac}}$;\\ {\bf let} $\zeta = \omega^{\star}_{\textsc{gs}}$, call \eqref{eqn11c} to compute $\mathbf{d}_{1}$; $t \leftarrow t + 1$;
			\STATE {\bf while}  $t \leq T_{\textsc{a}}$;\\ call \eqref{eqn11} to compute $\mathbf{s}_{t}$, ${\mathbf{x}}_{t}$, and $\mathbf{d}_{t}$; $t \leftarrow t + 1$;
			\STATE {\bf let} $\boldsymbol{\Theta} = \mathbf{D}^{-1}$; $\zeta = \omega^{\star}_{\text{Jac}}$;
			\STATE {\bf while} $t \leq T_\textsc{a}+T_\textsc{b}$; \\ 
			call \eqref{eqn11} to compute $\mathbf{s}_{t}$, ${\mathbf{x}}_{t}$, and $\mathbf{d}_{t}$; $t \leftarrow t + 1$; 
			\renewcommand{\algorithmicensure}{\textbf{END}}
			\ENSURE
		\end{algorithmic} }
\vspace{-0em}		
\end{algorithm}

The damping factors (i.e., $\omega^{\star}_{\textsc{gs}}$ and $\omega^{\star}_{\text{Jac}}$) can be computed using \eqref{eqn09}, respectively.
When implementing this algorithm, ANPID (GS) is able to offer fast convergence and close-to-MLSD performance simultaneously, as illustrated in \figref{fig01} with $T_{\textsc{a}} = 3$. 
By replacing $\mathbf{M}_{\textsc{gs}}$ with $\mathbf{M}_{\textsc{ssor}}$, the above algorithm can be adapted to become ANPID (SSOR).

\subsection{Complexity Analysis}
In this section, the computational complexity of the proposed methods is analyzed and compared with that of existing techniques.
The preprocessing of $\mathbf{A}$ and $\mathbf{b}$ is not considered, as it is common to all current MIMO detectors and can be parallelized.

For ANPID (GS), the serial complexity of computing $\mathbf{M}_{\textsc{gs}}^{-1}$ is $\mathcal{O}(N^2)$, owing to the triangular structure of $\mathbf{M}_{\textsc{gs}}$. 
The remaining calculations in ANPID (GS) can be executed in parallel.
The computation cost of $\mathbf{U}$ in the first iteration is $0.5N^2$, while calculating $\omega_{\textsc{gs}}$ and $\omega_{\text{Jac}}$ according to \eqref{eqn09} has a complexity of $MN+3M$.
In Stage A, the complexity of computing $\boldsymbol{\Theta}(\mathbf{b}-\mathbf{A}\mathbf{d}_{t-1})$ is $1.5N^2$, which reduces to $N^2+N$ in Stage B. 
The complexity of $\Gamma\left(\mathbf{s}_{t}\right)$ is negligible, as it involves no multiplication.
Finally, the complexity of damping process in \eqref{eqn11c} is $2N$.
As for ANPID (SSOR), the serial complexity of computing $\mathbf{M}^{-1}_\textsc{ssor}$ is also $\mathcal{O}(N^2)$, because $\mathbf{M}^{-1}_{\textsc{ssor}} =(\mathbf{M}^{H}_{\textsc{gs}})^{-1} \mathbf{D} \mathbf{M}^{-1}_{\textsc{gs}}$. 
The other components of ANPID (SSOR) are similar to those of ANPID (GS).

\begin{figure*}[t]
	\centering
	\subfigure{
		\begin{minipage}[t]{0.49\textwidth}	
			\label{fig02a}
			\centering
			\includegraphics[width=7.8cm]{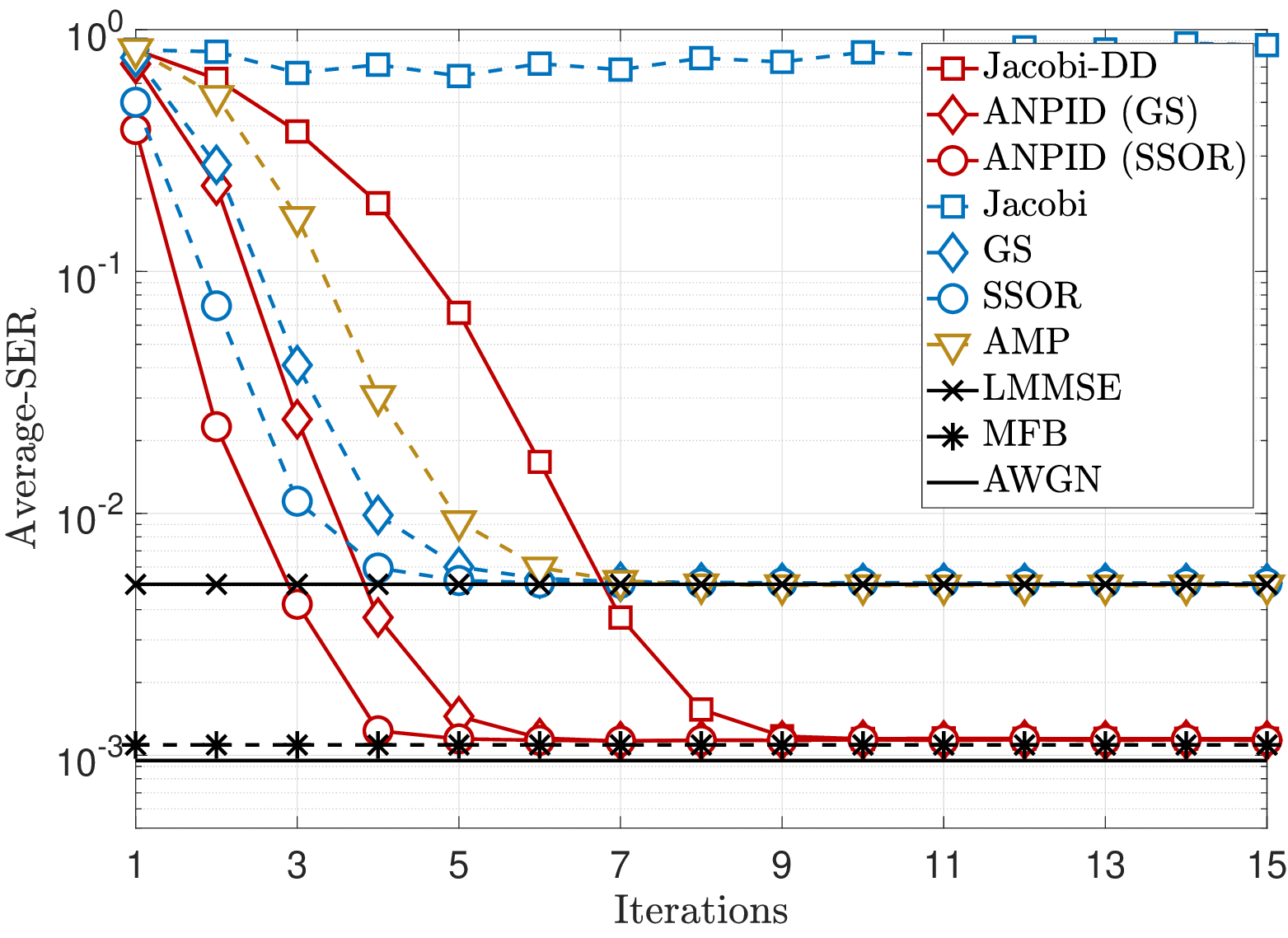}	
	\end{minipage}}
	\subfigure{
		\begin{minipage}[t]{0.49\textwidth}
			\label{fig02b}
			\centering
			\includegraphics[width=7.8cm]{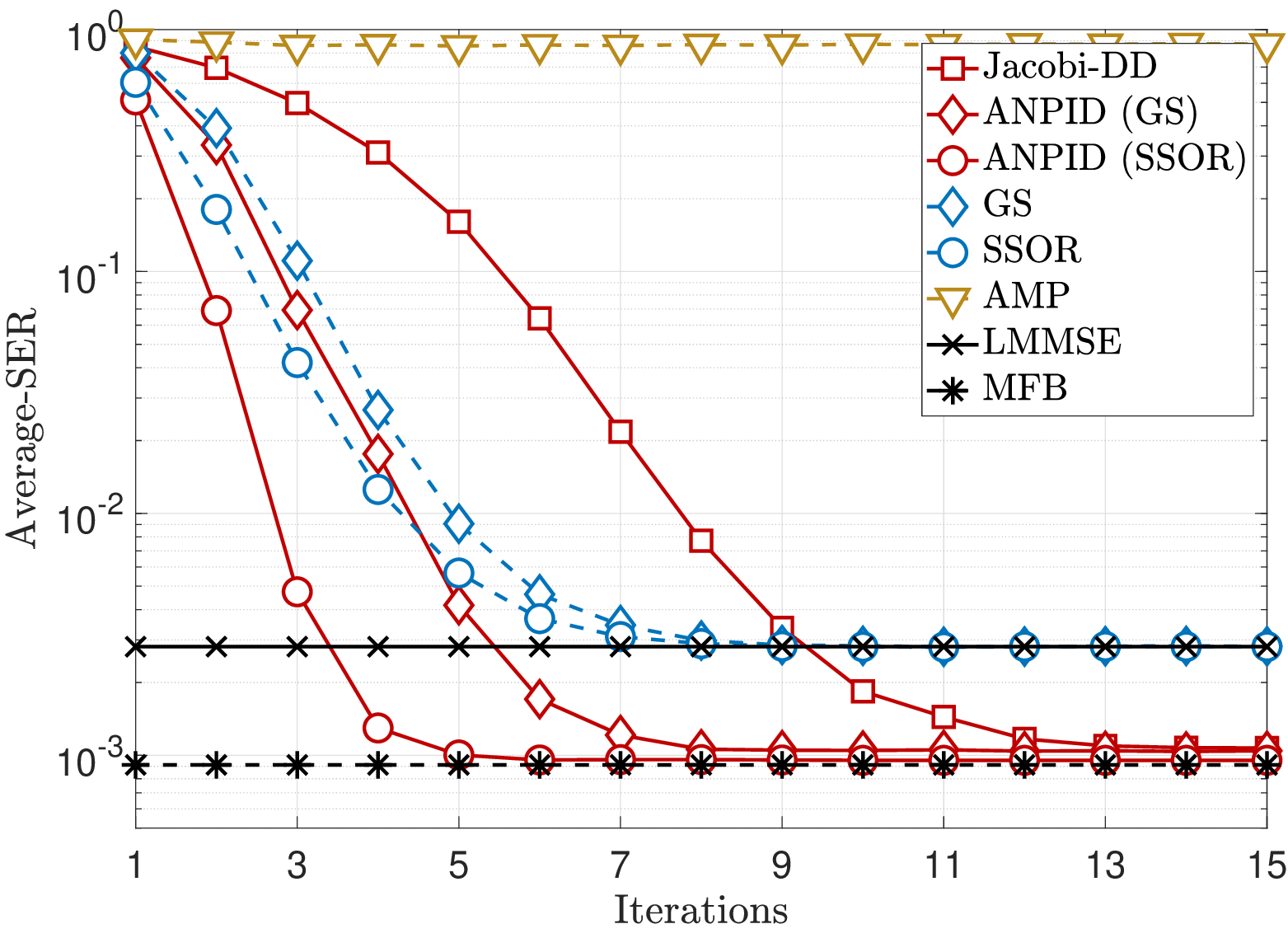}	   	
	\end{minipage}}		
	\caption{\label{fig02} SER versus iteration in $256 \times 64$ large-MIMO systems with $64$ QAM and $T_{\textsc{a}} = 3$. The three proposed methods can offer close-to-AWGN performance. {\bf Left:} WSSUS channel; Es/No = $24$ $\mathrm{dB}$. {\bf Right:} ELAA channel; Es/No = $31$ $\mathrm{dB}$.}
	\vspace{-1em}
\end{figure*}

\begin{table}[]
	\caption{Comparison of complexity and performance for different MIMO detectors}
	\label{tab01}
	\centering
	\renewcommand{\arraystretch}{1.3}
	\resizebox{0.48\textwidth}{!}{
		\begin{tabular}{|l|c|cc|c|}
			\hline
			\multicolumn{1}{|c|}{\multirow{2}{*}{Algorithm}} & \multirow{2}{*}{\begin{tabular}[c]{@{}c@{}}Serial \\ Complexity\end{tabular}} & \multicolumn{2}{c|}{Parallel Complexity (Per Iteration)}                                          & \multirow{2}{*}{Best Performance}                     \\ \cline{3-4}
			\multicolumn{1}{|c|}{}                           &                                                                                 & \multicolumn{1}{c|}{$t=1$}         & $t\geq2$               &                                                       \\ \hline
			Jacobi Iteration                                 & 0                                                                               & \multicolumn{1}{c|}{$N$}                     & $N^2$                          & LMMSE \\ \hline
			GS Iteration                                     & $N^2$                                                              & \multicolumn{1}{c|}{$0.5N^2$}                & $1.5N^2$                           & LMMSE \\ \hline
			SSOR                                             & $N^2$                                                              & \multicolumn{1}{c|}{$N^2 $}               & $2N^2$                         & LMMSE \\ \hline
			AMP                                              & 0                                                                               & \multicolumn{1}{c|}{$2MN$}                   & $2MN$                         &LMMSE                      \\ \hline
			PS-ADMM                                          & $N^3$                                                              & \multicolumn{1}{c|}{$MN$}                    & $N^2$               & Better than LMMSE                                        \\ \hline
			Jacobi-DD                                        & 0                                                                               & \multicolumn{1}{c|}{$2MN$}         & $N^2 $                         & Close-to-MLSD                                          \\ \hline
			ANPID (GS)                                       & $N^2$                                                              & \multicolumn{1}{c|}{$0.5N^2 + 2MN$} & $1.5N^2 \rightarrow N^2$ & Close-to-MLSD                                              \\ \hline
			ANPID (SSOR)                                     & $N^2$                                                              & \multicolumn{1}{c|}{$N^2 + 2MN$}    & $2N^2 \rightarrow N^2$   & Close-to-MLSD                                              \\ \hline
	\end{tabular}}
	\vspace{-1.5em}
\end{table}
\tabref{tab01} presents a comparison between the proposed MIMO detectors and existing techniques in terms of both complexity and performance.
To conserve space, the linear complexities have been excluded from the table since they have a negligible impact on the overall complexity of the MIMO detectors.
The table shows that existing MIMO detectors with square-order complexities can only provide LMMSE performance, while nonlinear MIMO detectors such as PS-ADMM can offer better than LMMSE performance, but at a cubic-order complexity.
In contrast, only the proposed methods (i.e., Jacobi-DD, ANPID (GS), and ANPID (SSOR)) can achieve close-to-MLSD performance with square-order complexities.
Jacobi-DD is capable of fully parallel computation, while ANPID (GS) and ANPID (SSOR) exhibit faster convergence rates among these methods.

\section{Simulation Results}
In this section, computer simulations are conducted in both WSSUS and ELAA channels.
To ensure a fair comparison, we set $\mathbf{x}_{0} = \mathbf{0}$ and $\mathbf{d}_{0}= \mathbf{0}$ for all the methods.
It is computational prohibitive to perform MLSD with large $N$ in Monte Carlo simulations.
Therefore, we use the performance of AWGN channel as the lower bound for WSSUS channel, and the matched filter bound (MFB) as the performance lower-bound for ELAA channels.
Denote $\mathbf{x}_{\textsc{mfb}}$ as the decision of MFB, it can be expressed as follows \cite{Amiri2022}
\begin{equation} \label{eqn04}
\mathbf{x}_{\textsc{mfb}} = \Gamma(\mathbf{s} + \mathbf{D}^{-1}\mathbf{v}),
\end{equation}
where $\mathbf{D}$ denotes the diagonal part of $[\mathbf{H}^{H}\mathbf{H}]$.
In MFB, it is assumed that the interference is perfectly eliminated, and then maximum ratio combining is used for each interference-free data stream. 
Therefore, MFB can serve as a performance lower bound for MIMO detectors.
For large-MIMO systems deployed with an ELAA, the service-antennas are deployed in a large uniform linear array (ULA) with half-wavelength equal spacing at a central frequency of $3.5$ $\mathrm{GHz}$.
The perpendicular distance between the ELAA and the users is set to be $15$ meters. 
According to \cite{Liu2021}, the system parameters are set as follows: $\alpha = 0.020$, $\beta = 1.765$, and $\sigma_s = 6.053$.
Three experiments are conducted in this section.

{\it Experiment 1:} The objective of this experiment is to evaluate the detection performance and convergence rate of various methods.
In \figref{fig02}, we present the SER versus iteration for both WSSUS and ELAA channels.
There is a considerable performance gap between LMMSE and MFB in both channels, which implies that all the linear MIMO detectors can only offer sub-optimal performance.
In WSSUS channels, the original Jacobi iteration fails, since the ratio of $N/M$ is too large \cite{Wang2022a}.
When comparing ANPID (GS) to GS or AUPID (SSOR) to SSOR, it becomes apparent that the former has a faster convergence rate. 
This is because ANPID methods incorporate a damping demodulation step into SI methods, which helps to improve convergence speed. 
As shown in the figure, only the three proposed methods (i.e., Jacobi-DD, ANPID (GS), and ANPID (SSOR)) are capable of achieving close-to-MLSD performance in both WSSUS and ELAA channels.
It is observed that ANPID (SSOR) has the fastest convergence rate, whereas Jacobi-DD takes advantage of its ability to support fully parallel computation despite having a slower convergence rate.
In the ELAA channel, all iterative methods exhibit slower convergence rates compared to that in WSSUS channels, due to the channels spatial non-stationarity.
Furthermore, it is evident that the AMP algorithm fails due to the channel spatial non-stationarity, while the proposed methods remain robust to ELAA channels and still offer close-to-MLSD performance.
Remarkably, for ANPID (SSOR) methods, very similar convergence rate (i.e., converging at $t=5$) is achieved in both WSSUS and ELAA channels.

\begin{figure*}[t]
	\centering
	\subfigure{\begin{minipage}[t]{0.49\textwidth}	
			\label{fig03a}
			\centering
			\includegraphics[width=7.8cm]{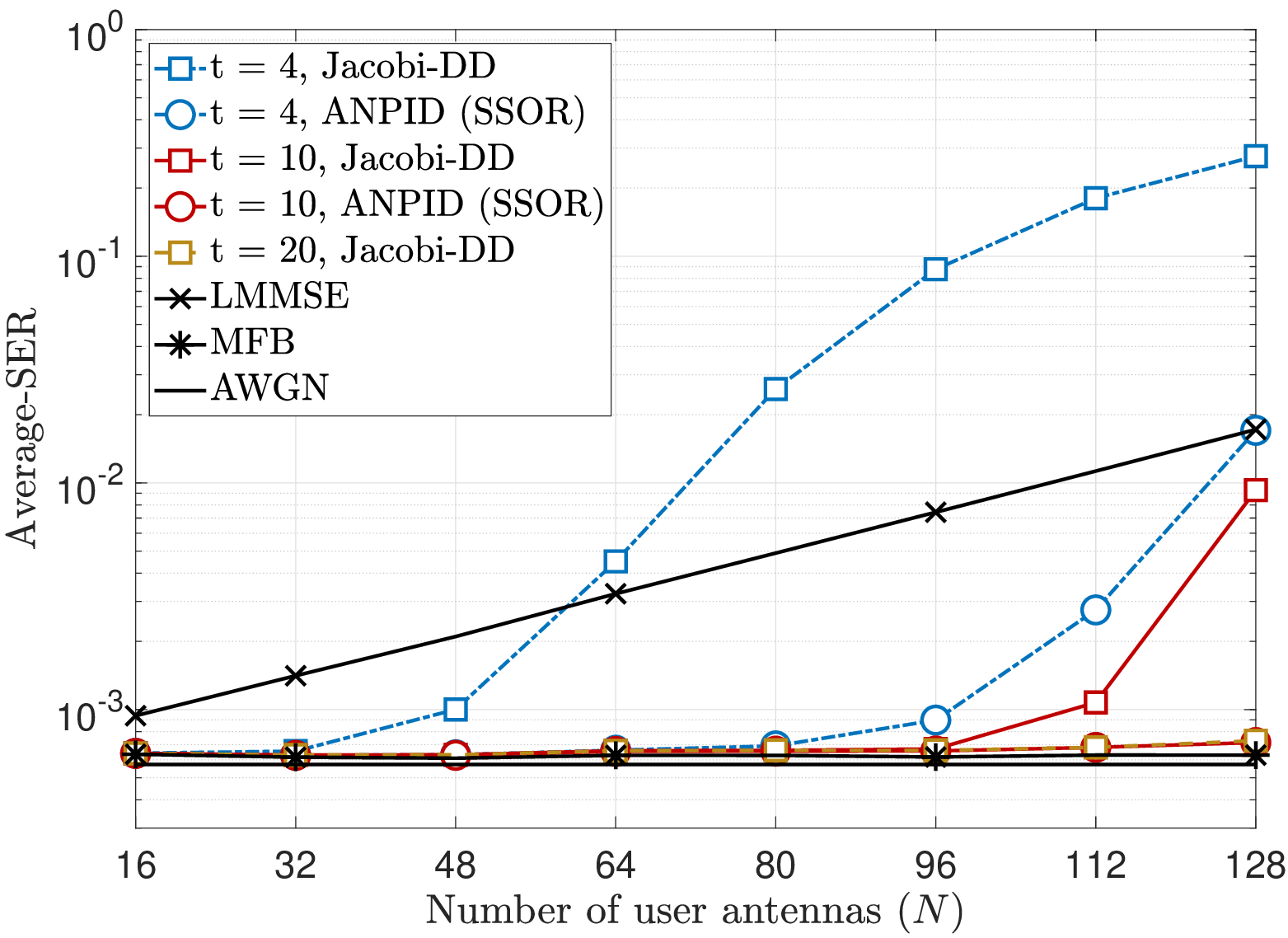}	
	\end{minipage}}
	\subfigure{\begin{minipage}[t]{0.49\textwidth}	
			\label{fig03b}
			\centering
			\includegraphics[width=7.8cm]{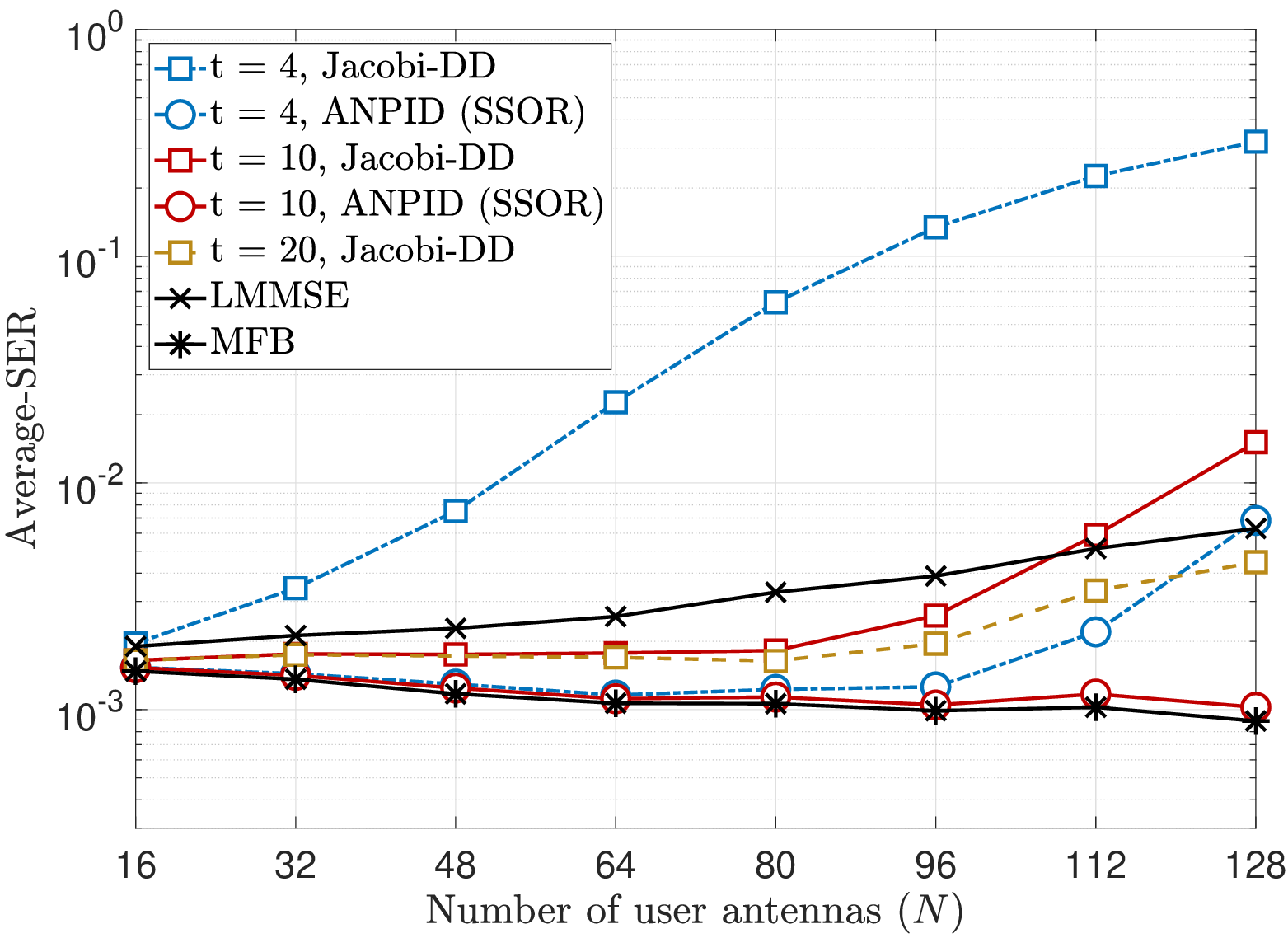}	
	\end{minipage}}
	\caption{\label{fig03} SER versus $N$ with $M$ fixed at $256$ using $16$ QAM modulation; $T_{\textsc{a}} = 3 $. It is demonstrated that the detection performance of the proposed methods are scalable as $N$ increases. {\bf Left:} WSSUS channel; Es/No = $18$ $\mathrm{dB}$. {\bf Right:} ELAA channel; Es/No = $27$ $\mathrm{dB}$.}
	\vspace{-1em}
\end{figure*}   

\begin{figure*}[t]
 	\centering
 	\subfigure{\begin{minipage}[t]{0.49\textwidth}	
 			\label{fig04a}
 			\centering
 			\includegraphics[width=7.8cm]{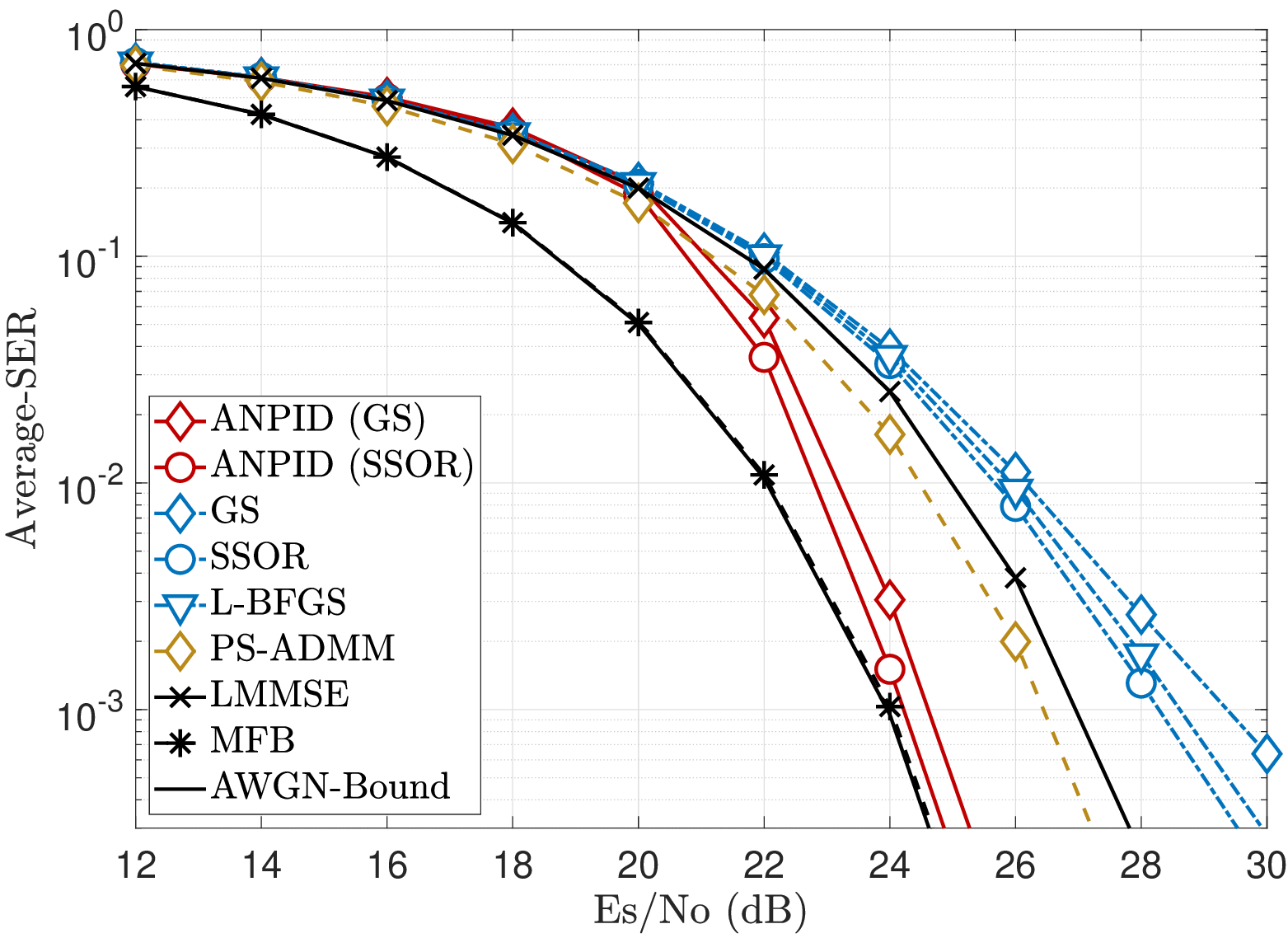}	
 		\end{minipage}}
 			\subfigure{\begin{minipage}[t]{0.49\textwidth}	
 					\label{fig04b}
 					\centering
 					\includegraphics[width=7.8cm]{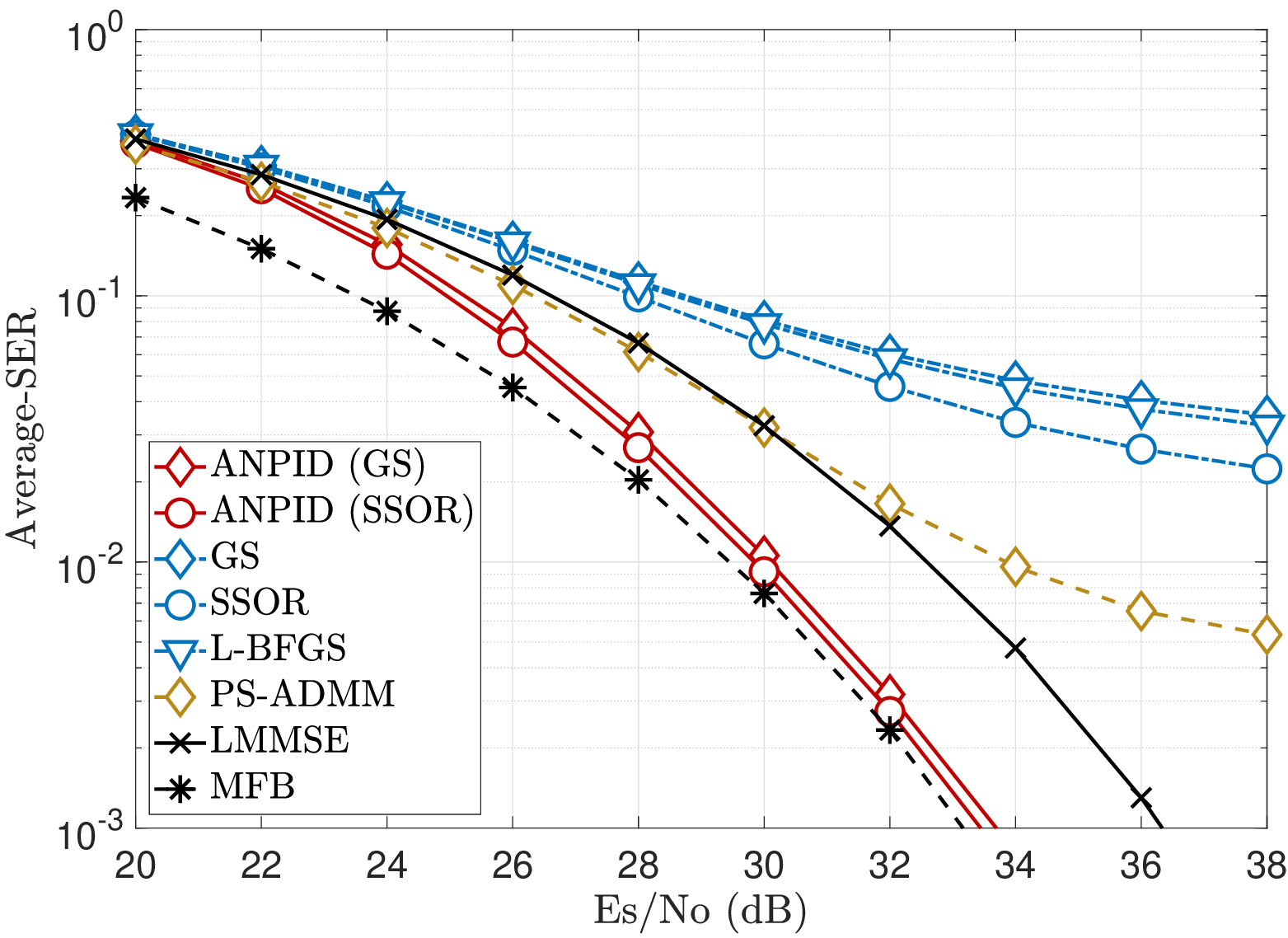}	
 				\end{minipage}}
 	\caption{\label{fig04} SER versus Es/No in highly-loaded large-MIMO systems with $M = 256$ and $N =128$; $64$ QAM; $T_{\textsc{a}} = 5$. The proposed ANPID methods can provide a performance gain of around $3$ dB when the SER is approximately $10^{-3}$. {\bf Left:} WSSUS channel. {\bf Right:} ELAA channel.}
 		\vspace{-1em}
\end{figure*}

{\it Experiment 2:} This experiment aims to demonstrate the scalability of the proposed methods in terms of detection performance as $N$ increases, while $M$ is fixed.
The results in \figref{fig03} show that ANPID (SSOR) and Jacobi-DD both provide close-to-AWGN performance in WSSUS channels.
This indicates that the performance using the proposed methods remains almost constant as the system load increases.
In ELAA channels, Jacobi-DD does not offer close-to-MFB performance (while AUPID (SSOR) does), but it still outperforms LMMSE.
The increasing inter-symbol interference in highly-loaded large-MIMO systems causes the performance gap between LMMSE and AWGN to widen. 
This implies that all low-complexity algorithms that can only offer LMMSE detection performance will be too sub-optimal in such systems.
In contrast, the results in \figref{fig03} indicate that ANPID methods can achieve close-to-MLSD while maintaining low computational complexity and fast convergence, which is crucial for practical large-MIMO systems.

{\it Experiment 3:} The purpose of this experiment is to evaluate the detection performance of the proposed methods in different levels of Es/No.
As shown in \figref{fig04}, the SER versus Es/No is plotted for various MIMO detectors in both WSSUS and ELAA channels.
The ANPID methods, compared to LMMSE, can provide a performance gain of about $3$ $\mathrm{dB}$ in both stationary and non-stationary MIMO channels, when SER is at $10^{-3}$.
It is notable that the performance gain between ANPID methods and LMMSE is greater at high $\mathrm{Es/No}$ ranges, which is typically the working range of large-MIMO systems.
In addition, the results show that PS-ADMM can only offer better-than-LMMSE performance in WSSUS channels.
However, PS-ADMM is no longer able to outperform LMMSE in ELAA channels.
On the other hand, ANPID methods demonstrate robustness to the spatial non-stationarity of ELAA channels.

\section{Conclusion}
In this paper, three iterative algorithms for large-MIMO detection were developed: Jacobi-DD, ANPUD (GS), and ANPID (SSOR).
The ANPID methods were shown to have faster convergence than Jacobi-DD, albeit with slightly higher complexity, and all methods can achieve close-to-MLSD performance with square-order complexity.
The simulation results showed that the proposed ANPID methods outperformed LMMSE by approximately $3$ $\mathrm{dB}$ in highly-loaded large-MIMO systems, which indicates that half of the transmit power can be saved in the physical layer.
The scalability and performance of the proposed methods make them suitable for practical use in modern wireless communication systems.
Finally, it is worth mentioning that the proposed methods are not limited to large-MIMO detection only, but can also be utilized to solve integer least squares problems in other domains, such as signal processing and computer vision.

\appendices
\section{Proof of {\it Theorem} \ref{thm01}} \label{app01}
Similar to the objective of MLSD in \eqref{eqn03}, it is proposed to calculated ${\omega}_{t}$ by minimizing the Euclidean distance between $\mathbf{y}$ and $[\mathbf{H}\mathbf{d}_{t}]$ as follows
\begin{equation}\label{eqn23591104}
	{\omega}_{t}  = \underset{\omega_{t}}{\arg \min }\|\mathbf{y}-\mathbf{H} \mathbf{d}_{t}\|^{2}.
\end{equation}
Plugging \eqref{eqn08c} into \eqref{eqn23591104} yields 
\begin{IEEEeqnarray}{ll}
	{\omega}_{t}\ &= \underset{\omega_{t}}{\arg \min }\|\mathbf{y}-\mathbf{H} \mathbf{x}_{t} - \omega_{t}(\mathbf{H}\mathbf{d}_{t-1} - \mathbf{H} \mathbf{x}_{t})\|^{2},  \nonumber \\ 
	&= \underset{\omega_{t}}{\arg \min }\|\boldsymbol{\tau}_{t} - \omega_{t}\boldsymbol{\nu}_{t}\|^{2},  \label{eqn14}
\end{IEEEeqnarray}
which is a quadratic function on $\omega_{t}$. 
Let $\frac{\partial}{\partial \omega}\|\boldsymbol{\tau}_{t} - \omega_{t}\boldsymbol{\nu}_{t}\|^{2} = 0$, \eqref{eqn15} in {\it Theorem \ref{thm01}} can be obtained.

\section{Prove of Theorem \ref{thm02}} \label{app02}
Eqn. \eqref{eqn10} in {\it Theorem \ref{thm02}} can be reformulated as follows
\begin{equation}\label{eqn16461204}
	\mathbf{s}_{t} = (\mathbf{I} - (\mathbf{M}\mathbf{U})^{-1} \mathbf{A}) \mathbf{d}_{t-1} + (\mathbf{M}\mathbf{U})^{-1}\mathbf{b}.
\end{equation}
Define $\mathbf{F} \triangleq (\mathbf{M}\mathbf{U})^{-1}\mathbf{A}$,
and plugging $\mathbf{b} = \mathbf{H}^{H}\mathbf{y}$ and $\mathbf{y} = \mathbf{H}\mathbf{x} + \mathbf{v}$ into \eqref{eqn16461204} yields
\begin{IEEEeqnarray}{ll}
	\mathbf{s}_{t}\ &= (\mathbf{I} - \mathbf{F})\mathbf{d}_{t-1} + (\mathbf{M}\mathbf{U})^{-1}\mathbf{H}^{H}(\mathbf{H}\mathbf{x} + \mathbf{v}), \nonumber \\
	&= \mathbf{F}\mathbf{x} + (\mathbf{I} - \mathbf{F})\mathbf{d}_{t-1} +  (\mathbf{M}\mathbf{U})^{-1}\mathbf{H}^{H}\mathbf{v}, \label{eqn10550304}
\end{IEEEeqnarray}
where the first two terms can be reformulated as follows
\begin{IEEEeqnarray}{ll}
 \mathbf{F}\mathbf{x} + (\mathbf{I} - \mathbf{F})\mathbf{d}_{t-1}\ &= \mathbf{F}\mathbf{x} - \mathbf{x} + \mathbf{x} + (\mathbf{I} - \mathbf{F})\mathbf{d}_{t-1}, \nonumber \\
 &= \mathbf{x} + (\mathbf{I} - \mathbf{F})(\mathbf{d}_{t-1}- \mathbf{x}). \label{eqn17111204}
\end{IEEEeqnarray}
Plugging \eqref{eqn17111204} into \eqref{eqn10550304} yields
\begin{equation}
	\mathbf{s}_{t} = \mathbf{x} + (\mathbf{I} - \mathbf{F})(\mathbf{d}_{t-1}- \mathbf{x}) + (\mathbf{M}\mathbf{U})^{-1}\mathbf{H}^{H}\mathbf{v},
\end{equation}
where the first term is the normalized signal vector.
From the definitions of $\mathbf{F}$ and $\mathbf{U}$, it is clear that the diagonal elements of $\mathbf{F}$ are all equal to $1$.
Consequently, $[\mathbf{I} - \mathbf{F}]$ is a hollow matrix, meaning the second term in \eqref{eqn17111204} contains only inter-symbol interference.
Furthermore, the third term corresponds to the noise vector and does not contain any signal component.
This confirms the validity of the expression derived in {\it Theorem \ref{thm02}}.

\section*{Acknowledgement}
This work is partially funded by the 5G Innovation Centre and 6G Innovation Centre.

\ifCLASSOPTIONcaptionsoff
\newpage
\fi

\bibliographystyle{IEEEtran}
\bibliography{../IEEEabrv,../mMIMO} 

\begin{thebibliography}{10}
\providecommand{\url}[1]{#1}
\csname url@samestyle\endcsname
\providecommand{\newblock}{\relax}
\providecommand{\bibinfo}[2]{#2}
\providecommand{\BIBentrySTDinterwordspacing}{\spaceskip=0pt\relax}
\providecommand{\BIBentryALTinterwordstretchfactor}{4}
\providecommand{\BIBentryALTinterwordspacing}{\spaceskip=\fontdimen2\font plus
\BIBentryALTinterwordstretchfactor\fontdimen3\font minus
  \fontdimen4\font\relax}
\providecommand{\BIBforeignlanguage}[2]{{%
\expandafter\ifx\csname l@#1\endcsname\relax
\typeout{** WARNING: IEEEtran.bst: No hyphenation pattern has been}%
\typeout{** loaded for the language `#1'. Using the pattern for}%
\typeout{** the default language instead.}%
\else
\language=\csname l@#1\endcsname
\fi
#2}}
\providecommand{\BIBdecl}{\relax}
\BIBdecl

\bibitem{Cui2023}
M.~Cui, Z.~Wu, Y.~Lu, X.~Wei, and L.~Dai, ``Near-field {MIMO} communications
  for {6G}: {Fundamentals}, challenges, potentials, and future directions,''
  \emph{{IEEE} Commun. Mag.}, vol.~61, no.~1, pp. 40--46, Jan. 2023.

\bibitem{BJORNSON20193}
E.~{Bj\"ornson}, L.~{Sanguinetti}, H.~{Wymeersch}, J.~{Hoydis}, and T.~L.
  {Marzetta}, ``Massive {MIMO} is a reality–{What} is next? {Five} promising
  research directions for antenna arrays,'' \emph{Digit. Signal Process.},
  vol.~94, pp. 3--20, Nov. 2019.

\bibitem{Liu2021}
J.~Liu, Y.~Ma, J.~Wang, N.~Yi, R.~Tafazolli, S.~Xue, and F.~Wang, ``A
  non-stationary channel model with correlated {NLoS/LoS} states for
  {ELAA-mMIMO},'' in \emph{Proc. IEEE Global Commun. Conf. (GLOBECOM)}, 2021,
  pp. 1--6.

\bibitem{Yang2015}
S.~Yang and L.~Hanzo, ``Fifty years of {MIMO} detection: The road to
  large-scale {MIMOs},'' \emph{{IEEE} Commun. Surveys Tuts.}, vol.~1, no.~4,
  pp. 1941--1988, 4th Quart. 2015.

\bibitem{Albreem2019}
M.~A. Albreem, M.~Juntti, and S.~Shahabuddin, ``Massive {MIMO} detection
  techniques: A survey,'' \emph{{IEEE} Commun. Surveys Tuts.}, vol.~21, no.~4,
  pp. 3109--3132, 4th Quart. 2019.

\bibitem{Chen2018}
Z.~Chen and E.~Björnson, ``Channel hardening and favorable propagation in
  cell-free massive {MIMO} with stochastic geometry,'' \emph{{IEEE} Trans.
  Commun.}, vol.~66, no.~11, pp. 5205--5219, Nov. 2018.

\bibitem{Liu2023a}
{J.~Liu, Y.~Ma, and R.~Tafazolli}, ``Leveraging user-wise {SVD} for accelerated
  convergence in iterative {ELAA-MIMO} detections,'' in \emph{Proc. IEEE 24th
  Int. Workshop Signal Process. Advances Wireless Commun. (SPAWC)}, 2023.

\bibitem{Albreem2022}
M.~A. Albreem, A.~H. Alhabbash, S.~Shahabuddin, and M.~Juntti, ``Deep learning
  for massive {MIMO} uplink detectors,'' \emph{{IEEE} Commun. Surveys Tuts.},
  vol.~24, no.~1, pp. 741--766, 1st Quart. 2022.

\bibitem{Shahabuddin2021}
S.~Shahabuddin, I.~Hautala, M.~Juntti, and C.~Studer, ``{ADMM}-based
  infinity-norm detection for massive {MIMO}: Algorithm and {VLSI}
  architecture,'' \emph{{IEEE} Trans. {VLSI} Syst.}, vol.~29, no.~4, pp.
  747--759, Feb. 2021.

\bibitem{Tiba2021}
I.~N. Tiba, Q.~Zhang, J.~Jiang, and Y.~Wang, ``A low-complexity {ADMM}-based
  massive {MIMO} detectors via deep neural networks,'' in \emph{Proc. IEEE Int.
  Conf. Acoust. Speech Signal Process. (ICASSP)}, 2021, pp. 4930--4934.

\bibitem{Zhang2022}
Q.~Zhang, J.~Wang, and Y.~Wang, ``Efficient {QAM} signal detector for massive
  {MIMO} systems via {PS/DPS-ADMM} approaches,'' \emph{{IEEE} Trans. Wireless
  Commun.}, vol.~21, no.~10, pp. 8859--8871, Oct. 2022.

\bibitem{Chen2017}
J.-C. Chen, ``A low complexity data detection algorithm for uplink multiuser
  massive {MIMO} systems,'' \emph{{IEEE} J. Sel. Areas Commun.}, vol.~35,
  no.~8, pp. 1701--1714, Jun. 2017.

\bibitem{Hochwald2004}
M.~Hochwald, T.~L. Marzetta, and V.~Tarokh, ``Multiple-antenna channel
  hardening and its implications for rate feedback and scheduling,'' \emph{IEEE
  Trans. Inf. Theory}, vol.~50, no.~9, pp. 1893--1909, Sep. 2004.

\bibitem{Liu2023}
J.~Liu, Y.~Ma, and R.~Tafazolli, ``Achieving maximum-likelihood detection
  performance with square-order complexity in large quasi-symmetric {MIMO}
  systems,'' in \emph{Proc. IEEE Int. Symp. Inf. Theory (ISIT)}, 2023.

\bibitem{Zhang2021}
C.~Zhang, Z.~Wu, C.~Studer, Z.~Zhang, and X.~You, ``Efficient soft-output
  {Gauss–Seidel} data detector for massive {MIMO} systems,'' \emph{{IEEE}
  Trans. Circuits Syst. {I}}, vol.~68, no.~12, pp. 5049--5060, Dec. 2021.

\bibitem{Xie2016}
T.~Xie, L.~Dai, X.~Gao, X.~Dai, and Y.~Zhao, ``Low-complexity {SSOR}-based
  precoding for massive {MIMO} systems,'' \emph{{IEEE} Commun. Lett.}, vol.~20,
  no.~4, pp. 744--747, Apr. 2016.

\bibitem{Amiri2022}
A.~Amiri, S.~Rezaie, C.~N. Manchón, and E.~de~Carvalho, ``Distributed receiver
  processing for extra-large {MIMO} arrays: A message passing approach,''
  \emph{{IEEE} Trans. Wireless Commun.}, vol.~21, no.~4, pp. 2654--2667, Apr.
  2022.

\bibitem{Wang2022a}
Z.~Wang, R.~M. Gower, Y.~Xia, L.~He, and Y.~Huang, ``Randomized iterative
  methods for low-complexity large-scale {MIMO} detection,'' \emph{{IEEE}
  Trans. Signal Process.}, vol.~70, pp. 2934--2949, Jun. 2022.

\end{thebibliography}
\end{document}